\documentclass[preprint,twocolumn,tightenlines,10pt,prb]{revtex4}%
\usepackage{amsfonts}
\usepackage{amsmath}
\usepackage{amssymb}
\usepackage{graphicx}%
\providecommand{\U}[1]{\protect\rule{.1in}{.1in}}

\begin{document}
\title{Signatures of Random Matrix Theory in the Discrete Energy Spectra of
Subnanosize Metallic Clusters}
\author{L. L. A. Adams$^{\ast}$, B. W. Lang, Yu Chen and A. M. Goldman}
\affiliation{School of Physics and Astronomy, University of Minnesota, 116 Church St. SE,
Minneapolis, MN 55455, USA}
\affiliation{$^{\ast}$Current address: The James Franck Institute, The University of
Chicago, 929 East 57$^{th}$ St., Chicago, IL 60637}

\pacs{PACS number}

\begin{abstract}
Lead clusters deposited on Si(111) substrates have been studied at low
temperatures using scanning tunneling microscopy and spectroscopy. The
current-voltage characteristics exhibit current peaks that are irregularly
spaced and varied in height. The statistics of the distribution of peak
heights and spacings are in agreement with random matrix theory for several
clusters. The distributions have also been studied as a function of cluster
shape. \TeX{} .

\end{abstract}
\volumeyear{year}
\volumenumber{number}
\issuenumber{number}
\received[Received text]{date}

\revised[Revised text]{}

\startpage{1}
\endpage{102}
\maketitle

\section{INTRODUCTION}

In the nearly 10 years that have passed since the first observation of
discrete energy levels in metallic clusters there still remains the question
of how the levels are statistically distributed in these systems. It has been
suggested that random matrix theory (RMT) \cite{RMT} is applicable to the
statistical properties of the spectra of metallic clusters in much the same
way that it is applicable to the slow neutron resonant spectra \cite{Garg}
observed in the 1950s and 1960s. However experimental verification of the
applicability of RMT in these systems is still challenging because of the
difficulty in gathering a sufficient number of levels to analyze their
statistical distribution.

The first observations of discrete energy levels or \textquotedblleft particle
in a box energy levels\textquotedblright\ in metallic clusters were made by
Ralph, Black and Tinkham \cite{Ralph} in 1995. The energy levels were observed
as irregular steps contained within the Coulomb staircase in the
current-voltage characteristics of clusters that were fabricated using a fixed
tunneling geometry with metallic electrodes. These uneven steps in the
current-voltage measurements might be a consequence of random matrix theory
(RMT). \cite{Narvaez} This suggestion arises from earlier predictions that
address these systems from various theoretical standpoints, including Efetov's
supersymmetry derivation.\cite{Efetov} While these expectations are
theoretically well established they are experimentally difficult to realize
because of non-equilibrium effects \cite{Agam} and capacitive charging energy
terms that have a tendency to mask the energy levels in mesoscopic systems.
Subsequent tunneling experiments have been performed on metallic clusters
\cite{Wang}\ and semiconducting dots \cite{InAs}$^{,}$\cite{CdSe} which have
yielded results similar to those of Ralph \textit{et al}., although the nature
of the level statistics still remains elusive.

In contrast to metallic clusters, experimental work has exhaustively addressed
the issue of distributions of level spacings and eigenfunctions of quantum
dots fabricated from two dimensional electron gas systems of various
pre-defined shapes.\cite{Sivan} In these systems, electron-electron
interactions dominate transport and the level spacing distributions appear to
be Gaussian, while the distributions of the amplitudes of the eigenfunctions
follow a Porter-Thomas distribution which is a signature of random matrix
theory (RMT).

Random matrix theory (RMT)\ and quantum chaos were merged in the conjecture
put forward by Bohigas, Giannoni and Schmit \cite{Bohigas} in 1984. This
conjecture states that the nearest neighbor energy level spacings of
classically chaotic systems should be distributed according the Gaussian
Orthogonal Ensemble (GOE), or Wigner-Dyson \cite{WD} distribution and this
conjecture is strongly supported by aggregated numerical studies.

The Wigner-Dyson distribution which describes the statistical distribution of
nearest neighbor energy levels normalized to the mean energy level has several
important features as described by Porter. \cite{PT} First, the probability of
having nearest neighbors with zero spacing disappears. Second, the probability
of a level spacing is linear in energy before approaching a maximum, with the
maximum occurring close to the mean energy and the tail of the distribution
being fairly small. This is in contrast with completely random levels (a
classically non-chaotic system) where the distribution is Poissonian. For the
latter distribution the probability is largest at zero level spacing. The
absence of small spacings in the Wigner-Dyson distribution is known as the:
"repulsion of energy levels". \cite{PT} This is the key ingredient of a
Wigner-Dyson distribution and distinguishes classically chaotic from
classically non-chaotic systems.

Besides the distributions of the eigenvalues which are most often addressed
experimentally, it is also possible to study the statistics of the amplitudes
of the eigenfunctions within the context of RMT. The corresponding
distribution is called the Porter-Thomas \cite{PT} distribution. The
Porter-Thomas distribution is simply a statement about the amplitude $\psi$ of
a wavefunction at any given point is a random variable. The distribution of
the square of a random variable, $\left\vert \psi\right\vert ^{2},$which is
Gaussian distributed, is the Porter-Thomas distribution.

It is also interesting to note that it is possible, in some systems, to tune a
transition from regular to chaotic behavior in which the distributions of the
nearest neighbor level spacing change from Poisson to Wigner-Dyson. This has
been found numerically in the study of the Rydberg levels of the hydrogen atom
\cite{Monteiro} in which the transition is tuned by the application of an
increasingly strong magnetic field and experimentally in diamagnetic helium by
tuning the excitation energy in the presence of a magnetic
field.\cite{Karremans}

Here we present experimental results that indicate that statistical
distributions of highly irregularly shaped Pb clusters follow RMT. This is an
extension of earlier work \cite{Adams PRL} and its purpose is to address in
detail experimental issues and present additional results which led to our
conclusions. The present paper is divided into five major sections as follows:
in Section II we treat the device geometry that was configured to study the
discrete energy level spectra of metallic clusters. In Section III, we relate
energy spectra to the geometry and discuss in detail the observed features of
the spectra. In Section IV we discuss the results of the preceding section in
terms of statistical distributions of both the eigenvalues and the amplitude
of the moduli of the eigenfunctions. In Section V, we demonstrate the use of
scanning tunneling spectroscopy to resolve real time images of a quantity that
is proportional to the square of the amplitude of the eigenfunctions in these systems.

\section{DEVICE GEOMETRY}

Before describing the current-voltage characteristics, it is worthwhile to
summarize some of the salient features of the device configuration relevant to
this work that are different from other spectroscopy measurements on metallic
clusters. The tunneling geometry that was used is different from that used
previously in three important respects. First, Pb clusters were grown on a
semiconducting substrate. This substrate was highly resistive with a
resistivity at room temperature greater than 1000 $\Omega$\textperiodcentered
cm. \cite{Virginia} Second, the clusters were fabricated using the buffer
layer assisted growth technique that was developed by Weaver and co-workers
\cite{Weaver} in which Xe, an inert gas, is used as a buffer layer to control
the size distribution of the clusters. This physical technique allows the
clusters to land softly onto the substrate and involves no use of chemicals or
organics in the growth process. Third, fabrication of the clusters and
spectroscopic measurements were carried out \textit{in situ} in an ultrahigh
vacuum environment thus preventing contamination.%
\begin{figure}
[ptb]
\begin{center}
\includegraphics[
height=1.1978in,
width=1.6873in
]%
{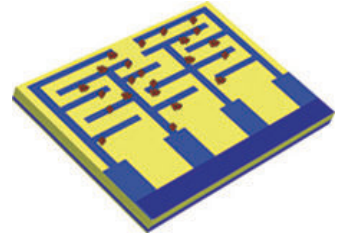}%
\caption{An illustrative drawing of the device geometry. \ Electrodes were
predeposited on the top of the Si substrate in the configuration shown in the
diagram and contacted to the sample plate. \ On the back side of the 500 $\mu
m$ thick Si substrate, a uniform bilayer of Ti/Pt was deposited for electrical
contact to the sample plate.}%
\label{fig1}%
\end{center}
\end{figure}

From the standpoint of conventional scanning tunneling microscopy (STM)
experiments it is also unique in both the use of a highly resistive substrate
and the addition of a bilayer of Ti/Pt electrodes on top of the substrate. The
device geometry is illustrated in Fig. \ref{fig1}. The first barrier
separating the STM tip (not shown) from the cluster is a vacuum barrier while
the second barrier between the cluster and the substrate is a thin silicon
oxide barrier. This double tunnel junction arrangement is necessary for the
resolution of discrete energy levels in these nanostructures. The experimental
details of the cluster fabrication have been published elsewhere. \cite{Adams}
The buffer layered assisted growth involves initially depositing four
monolayers of Xe onto a cold substrate that is held at temperatures less than
50K. The pressure of the Xe is controlled upon entry into the deposition
chamber by a capacitance manometer. The condensed Xe is then subjected to a
precisely controlled exposure of Pb vapor flux. The average film thickness of
Pb was less than 0.2 \AA \ as measured by a calibrated quartz crystal
oscillator. Subsequently the substrate is slowly warmed to room temperature
such that the Xe desorbs and the Pb clusters softly land onto the substrate.
The resulting sample which is held on a rotable liquid helium cooled transfer
rod is moved \textit{in situ} into the STM\ chamber through a large gate valve
and placed via a wobble stick onto the STM\ stage. All STM measurements were
made with the STM\ operating at 4.2 K although the electron temperature is
significantly larger. The STM\ tip that was used in the experiment was made
from tungsten wire. The tip was characterized prior to making spectroscopy
measurements by demonstrating atomic scale resolution of a graphite surface.

After fabrication, the Pb clusters were characteristized using STM. The
heights of the clusters were measured and found to be mostly between 3
\AA \ to 12 \AA , with an average cluster height of 8 \AA . When a bias
voltage was applied between the tip and counter-electrodes on top of the Si
substrate with the cluster in between, the current was measured through the
cluster. Current flows when the levels of the cluster come into juxtaposition
with the level(s) at the interface state or another set of discrete energy
levels from a neighboring cluster. The peaks observed in the current-voltage
curves (Fig. \ref{fig2}) are indicative that resonant tunneling processes are
operative whenever one of the cluster's quantized states is probed. This will
be further elaborated upon in the next section.

\section{CURRENT-VOLTAGE CHARACTERISTICS}

In this section we shall be concerned with measurements of current-voltage
characteristics of a series combination of two sets of discrete energy levels
of systems separated by\ insulating barriers. Electron transmission through
the clusters was probed by the STM\ tip in spectroscopy mode. The tunneling
current was recorded as a function of applied sample bias voltage $V$ while
the STM feedback loop was open. In this case, the tip was held at a fixed
height above the cluster while the current-voltage characteristics were
recorded. The data presented in this work were all obtained in voltage step
sizes of 480 $\mu V$ or 612 $\mu V$ in a time frame $\geq\,$0.2 $ms$ per
point.
\begin{figure}
[ptb]
\begin{center}
\includegraphics[
height=2.9291in,
width=3.0441in
]%
{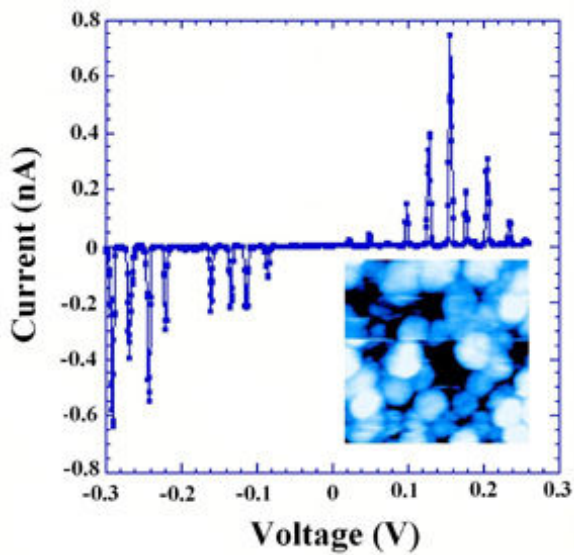}%
\caption{Tunneling current versus voltage at T = 4.2 K. \ Tunneling is from a
tunsten STM\ tip into a Pb cluster. Inset: 30.0 nm x 30.0 nm image of Pb
clusters grown by a buffer layer assisted growth technique. This image was
obtained using a bias voltage of -3.0 $V$ with a tunneling current of 2.0 x
10$^{-9}\,A.$}%
\label{fig2}%
\end{center}
\end{figure}

Due to resonant tunneling between the tip, the cluster, the interface state,
and finally the Pt electrode, the $I-V$ characteristics exhibit peaks as shown
in Fig. \ref{fig2}. These peaks are irregularly spaced and varied in height
and are dependent on cluster size. The statistics of peak heights and spacings
will be discussed in the next section (Sec. IV). Several hundred $I-V$ curves
were obtained on each cluster and their character varied along a given cluster
in a nonsystematic manner.

The magnitude of the differential conductance is an appropriate measure of
whether or not the tunneling regime is the correct description of the
transport. This will be addressed in this section along with a brief
discussion of elastic and inelastic tunneling as it pertains to the
linewidths. Also the relationship between the $I-V$ curves and the sizes of
the clusters will be discussed. But first we will take up the issue of
resonant tunneling in this system. The data is quite different from that which
would be obtained in a \ geometry in which metallic atoms (or clusters) are
deposited on an oxidized metallic surface and transmission is via nonresonant
process. \cite{Ralph}

\subsection{Resonant Tunneling Mechanism}

The observation of peaks instead of steps strongly suggests that the transport
through the cluster is due to resonant tunneling processes. \cite{Chang}$^{-}%
$\cite{Wilkinson} One possible explanation of how this occurs is the existence
of an interface state between the Pb clusters and the Si substrate. It is
known from photoemission experiments that such an interface exists between Pb
and Si(111) and this interface state is nearly dispersionless.\cite{pinning}
While there are differences in resistivity between the Si substrate that was
used in our study and the photoemission experiment, we assume that such a
state exists.

In all the $I-V$ spectra, the tunneling current is suppressed around zero bias
followed by a series of peaks at both negative and positive bias. The current
is suppressed when the energy levels of the cluster are not in registry with
the interface state. The double barrier tunnel junction that is realized by
positioning the STM tip over a Pb cluster is highly asymmetric with the first
tunneling barrier between the tip and the cluster and the second barrier
between the cluster and the substrate. This asymmetry is manifested in the
asymmetry of the $I-V$ curves about zero bias.

Resonance occurs when the Fermi wavelength spans the length of the cluster.
The Fermi wavelength for Pb is 4 \AA \ at room temperature.\cite{Kittel} Given
that the average height is 8 \AA , and measurements were carried out at liquid
helium temperatures, the assumption that the tunneling mechanism is a resonant
process is appropriate.

\subsection{Level Spacing as a function of Cluster Size}

Table \ref{Pb Parameters} catalogs the Pb clusters with the estimated number
of atoms per cluster, and estimated and measured mean level spacings. The
volume of the clusters was calculated assuming that each cluster is a
hemispherical cap such that the volume, $Vol$. is $\frac{\pi h}{6}%
(3r^{2}+h^{2})$, where $r$ is the radius and $h$ is the height of the cluster.
Since the clusters were irregularly shaped, their radius was estimated from
$r=\sqrt{\frac{lw}{\pi}}$ where $l$ is the length of the cluster and $w$ is
the width. The estimated number of atoms was calculated from $Vol./\left(
\frac{1}{4}a^{3}\right)  $ where $a$ is the lattice parameter of Pb and is
equal to 4.95 \AA .\cite{Kittel} The estimated mean level spacing, which was
calculated using the nearly free electron model, is $\left\langle
\Delta\right\rangle =\frac{2\pi^{2}\hbar^{2}}{m\,k_{f}\,Vol}$\ and compared to
the measured mean level spacings. Several of the clusters' measured mean level
spacing were in the range of $8-10\,meV$. There are several explanations for
this. First, the majority of the clusters had the same height which is a
result of the buffer layer assisted growth technique when using a four
monolayer thick buffer layer. Second, the calculation of the volume is a rough
estimate and not entirely accurate in that it oversimplifies the actual shape
of the clusters. Third, the clusters were not well isolated from each other
thus the width and the length could be larger if more than one cluster was
taken into consideration. Also, tip convolution effects obscure the actual
width and length of the clusters. Nonetheless there is an apparent dependency
on the level spacing energy on the size of the clusters.%

\begin{table*}[tbp] \centering
\begin{tabular}
[c]{|c|c|c|c|c|c|c|}\hline
Cluster & Length & $Vol$ &
$<$%
N%
$>$%
& Est.
$<$%
$\Delta$%
$>$%
& Meas.
$<$%
$\Delta$%
$>$%
& $\partial E\,\,\partial\tau_{SO}$\\
Index No. & (nm) & (nm)$^{3}$ &  & (meV) & (meV) & ($\hbar$ = 1)\\\hline
1 & 2.7 & 2.6 & 85 & 36.8 & 24.2 $\pm$ 5.39 & .083\\\hline
2 & 3.32 & 3.81 & 125 & 25.1 & 6.35 $\pm$ 0.74 & .070\\\hline
3 & 2.29 & 3.86 & 127 & 24.76 & 9.02 $\pm$ 1.45 & .047\\\hline
4 & 3.61 & 6.29 & 207 & 15.2 & 10.37 $\pm$ 0.43 & .046\\\hline
5 & 3.13 & 6.3 & 208 & 15.15 & 9.41 $\pm$ 1.12 & .04\\\hline
6 & 3.04 & 7.36 & 243 & 12.99 & 9.13 $\pm$ 0.29 & .033\\\hline
7 & 2.7 & 7.4 & 244 & 12.91 & 10.8 $\pm$ 0.54 & .029\\\hline
8 & 3.88 & 7.8 & 257 & 12.25 & 8.39 $\pm$ 0.28 & .04\\\hline
9 & 5.0 & 8.2 & 270 & 11.65 & 9.01 $\pm$ 0.26 & .049\\\hline
10 & 3.73 & 9.62 & 317 & 9.93 & 8.58 $\pm$ 0.34 & .031\\\hline
11 & 3.12 & 10 & 329 & 9.55 & 5.37 $\pm$ 1.53 & .025\\\hline
12 & 4.3 & 12.7 & 418 & 7.52 & 9.6 $\pm$ 1.25 & .027\\\hline
13 & 2.54 & 13.2 & 435 & 7.23 & 8.98 $\pm$ 1.63 & .0154\\\hline
14 & 5.49 & 19.4 & 638 & 4.94 & 7.35 $\pm$ 1.30 & .023\\\hline
15 & 4.12 & 21.5 & 709 & 4.44 & 7.49 $\pm$ 0.37 & .015\\\hline
16 & 4.76 & 27.1 & 892 & 3.53 & 4.73 $\pm$ .87 & .014\\\hline
\end{tabular}%
\caption{Summary of physical parameters of Pb clusters of different shapes and
sizes. \ As discussed in the text, the \textit{Vol.} is the volume of a
cluster and is estimated assuming it is a hemispherical cap. The number
of atoms,<N>, and the est. mean level spacing are calculated from this est. volume.
The est. mean level spacing is based on the nearly free electron model and
compared to experimentally measured mean level spacing values. The
expression $\partial E\,\partial\tau_{SO}%
$ is related to the cluster's size and is described in the text (see Sec. IV D).}%
\label{Pb Parameters}%
\end{table*}%

\subsection{Lineshapes and Differential Conductance}

Information about the intrinsic lifetime of the electronic states of a single
metal cluster and the tunneling process (inelastic or elastic) is in part
difficult to obtain because of the temperature of the reserviors. This can be
circumvented by having two sets of electronic states separated by a barrier
through which the current is measured. In this tunneling scenario, electron
transport is expected to occur only when the energy levels of the two states
are aligned with one another \cite{Nazarov} and the upper bounds of the
lifetimes of energy levels of varying size clusters can be estimated from the
average widths of the resonances that results from these level alignments. The
reason that these widths are upper bounds is that they depend on both the
reciprocals of the lifetimes of the electronic states and the tunneling rates.
We fit the full width half maximum of the peak widths in the I-V curves for
three clusters to Lorentzians (Lorentzians provided the best fit to the data)
and found that they varied depending on the cluster size. For cluster 1, the
average peak width was 3.54 meV (tunneling time: 1.9 $\times10^{-13}%
\,$seconds), for cluster 7, the average peak width was 3.06 meV (tunneling
time: \ 2.4 $\times10^{-13\,}\,$seconds) and for cluster 12, the average peak
width was 1.06 meV (tunneling time: 6.2 $\times10^{-13}\,\,$seconds)
indicating that it takes longer to tunnel through a larger cluster than a
smaller one. Lorentzian \ peak widths are also an indication that the
tunneling process is elastic. Moreover, the absence of diffusion (the
transport is ballistic) together with the spatial resolution of the
STM\ enables detailed studies of the clusters' electronic states.

\begin{figure}
[ptb]
\begin{center}
\includegraphics[
height=2.9162in,
width=3.0441in
]%
{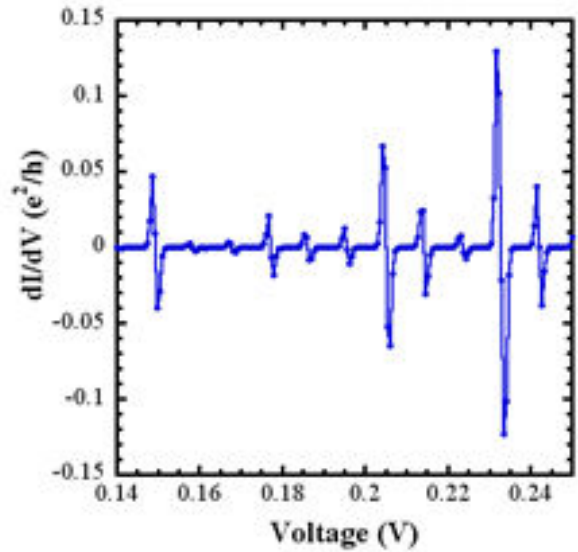}%
\caption{Differential conductance (in terms of e$^{2}/h$) versus voltage at T=
4.2K for cluster no. 12. The symmetry about the y-axis and the magnitude of
the peaks indicates that elastic tunneling processes are relevant.}%
\label{conductance}%
\end{center}
\end{figure}

Figure \ref{conductance} shows a plot of the differential conductance in units
of e$^{2}$/h plotted against positive bias for one of the clusters. This is
supporting evidence that the features are related to elastic tunneling
processes as the conductance is less than one and the peaks are Lorentzians.%

\subsection{Position Space Representation of Eigenstates}

\begin{figure}
[ptb]
\begin{center}
\includegraphics[
height=3.2206in,
width=2.7994in
]%
{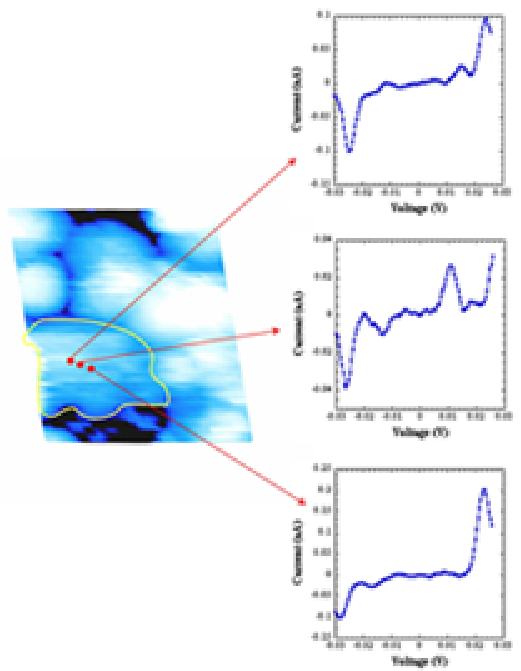}%
\caption{STM image (12 nm x 8.0 nm) of Pb clusters. The $I-V$ curves
correspond to three different positions along a cluster. There is not any
apparent correlation between the curves. The current setpoint = 1.0 nA.}%
\label{posdep}%
\end{center}
\end{figure}

An important feature of the measurements was that the current-voltage
characteristics varied with position on a cluster as shown in Fig.
\ref{posdep}. The peak heights, the number of peaks, and voltages where the
peaks occuried varied randomly as a function of position.%
This dependence is reminiscent of a quantum chaotic system in which the
amplitude of the eigenfunctions are known to vary in a complicated pattern.
\cite{Backer} There was no noticeable systematic variation in the curves as a
function of distance. It was not possible to reproduce the curves because of
the difficulty in repeating measurements at the exact same pixel point due to
drift in the piezos which control the tip's position.

\begin{figure}
[ptb]
\begin{center}
\includegraphics[
height=7.715in,
width=1.9614in
]%
{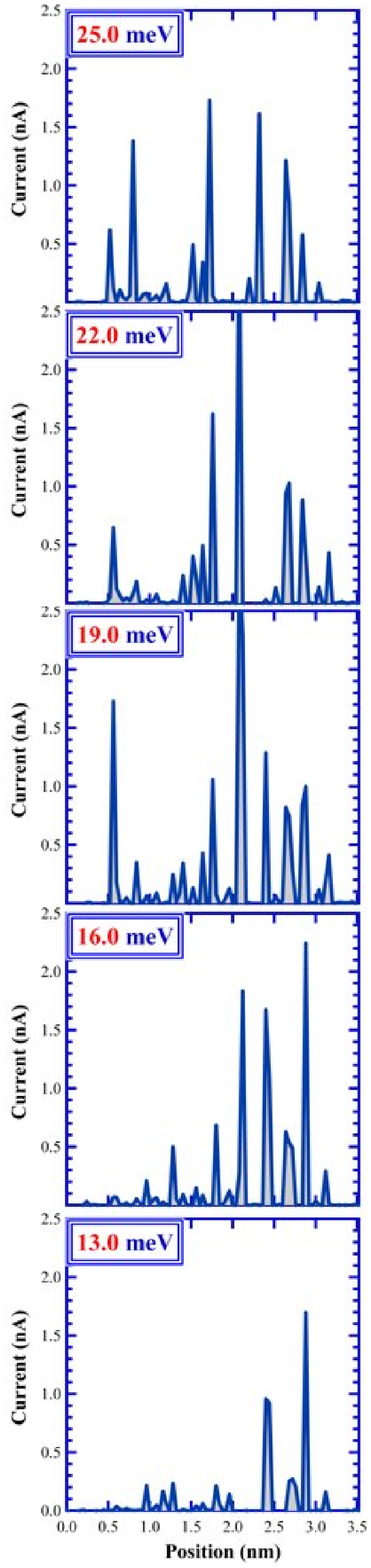}%
\caption{A map of the current as a function of position along a chaotic
cluster (catalog number 4) as the bias voltage is varied. \ There is noticable
clustering of the current peaks. \ }%
\label{cluster52}%
\end{center}
\end{figure}

For a given current profile across a cluster, a map of the current as a
function of the voltage bias can be obtained as shown in Fig. \ref{cluster52}.
For cluster no. 12, such a map displays strong clustering of current peaks as
a function of applied bias voltage although there is no obvious voltage
dependence.%

\subsection{Other possible tunneling scenarios}

The current voltage characteristics bare a strikingly similar resemblance to
the current vs \textit{gate} voltage curves seen in the results reported by
Kouwenhoven, Austing and Tarucha. \cite{Kouwenhoven} It might be the case,
that neighboring clusters are actually gating the cluster under investigation.
This would offer an alternative explanation to the position dependence of the
I-V curves as the capacitance will change due to the distance between the
gating cluster and the site on the cluster that was being probed by the STM
tip. Further work is needed to elucidate the mechanism behind the peaks in the
current-voltage characteristics. One such possible experiment would be to
fabricate the clusters without the electrodes configured on top of the
substrate. This would provide a more uniform distribution of clusters and
increase the separation between clusters since the clusters tended to
conglomerate near the electrodes.

\section{STATISTICAL DISTRIBUTIONS}

Current peaks in the scanning tunneling spectroscopy were not expected.
However, in an attempt to interpret the results the data was analyzed with the
assumption that the peaks were signatures of discrete energy levels. Exact
agreement with different sets of discrete energy levels for both the negative
and positive bias voltages would demand that the capacitances between both the
tip and cluster and cluster and substrate are substantially different, which
we believe is the case. Regardless, from the I-V measurements themselves, it
is not possible to confirm that the peaks in the negative and bias voltages
arise from different energy levels without further analysis. Therefore, based
on the premise that the discrete energy spectra in metallic clusters should
follow RMT, we studied the statistical distributions of both the peak spacings
and peak heights. Peaks in the I-V characteristics were identified after
running a smoothing program \cite{footnote} twice through the data. This
program was used to identify all the peaks. Since the voltage range over which
peaks were found in this study was small (-35 meV to +30 meV), the likelihood
of missing any peaks was eliminated. A practical matter connected with the
reliability of the histograms must also be mentioned. The number of bins used
in the histograms was the range divided by the experimental step size. The
histograms were rescaled using the measured mean level spacing to normalize
the mean to unity. The fits to the resulting histogram were carried out using
MINOS\ \cite{root} which is a minimization algorithm implemented in MINUIT.
\cite{MINUIT} The parameters that are obtained are those that correspond to a
minimum chi-squared value. Also, the peak widths are approximately ten times
greater than the bin size, thus the peaks spacing distribution that was
generated is not due to a "binning" effect.

\subsection{Wigner-Dyson Statistics}%

\begin{figure}
[ptb]
\begin{center}
\includegraphics[
height=1.7884in,
width=2.4249in
]%
{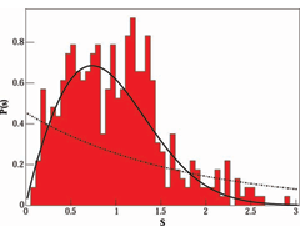}%
\caption{Histogram of peak spacings for cluster no. 9. The solid curve is the
fit for the Wigner-Dyson distribution. The dotted line represents the fit for
the Poisson distribution. There are 413 peak spacings that comprise this
histogram normalized to the mean voltage spacing.}%
\label{WD}%
\end{center}
\end{figure}

The histogram of normalized peak spacings (normalized to the mean spacing of
each individual trace) was fit by the distribution function%
\begin{equation}
P(s)\,=\,b_{\beta}\,s^{\beta}\,\exp(-a_{\beta}\,s^{c_{\beta}}) \label{Wigner}%
\end{equation}
The distribution function P(s) is a probability density defined such that the
area under the curve is one.\textit{ }Here the normalized mean spacing, $s$,
is simply $\Delta\,/\,\langle\Delta\rangle\,$, with $\Delta$ representing the
level spacing and $\langle\Delta\rangle$ the mean spacing. Equation
\ref{Wigner} \ can represent the orthogonal ($\beta\,=\,1$), unitary
($\beta\,=\,2$) and symplectic ($\beta\,=\,4$) ensembles \cite{Wigner-Dyson}
that correspond to processes with different symmetries. The orthogonal case
corresponds to time reversal symmetry being preserved in the absence of a
magnetic field and describes the results presented here.\ In the statistical
analysis of this histogram fits by Wigner-Dyson, Poisson, Gaussian, and
Lorentzian distributions were made. From the values of $\chi^{2}$(not shown)
it is clear that the Wigner-Dyson distribution provides the best fit to the
data with $a_{\beta}\,=\pi/4,b_{\beta}=\pi/2$ and $c_{\beta}=2$. \ In
Fig.\thinspace\ref{WD}, the histogram of peak spacings for this cluster,
showing the Wigner-Dyson and Poisson fits is plotted. (In the figure, the
fitted Poisson distribution is not one about the origin since the parameters
were allowed to float in order to minimize chi\symbol{94}2).

\subsection{Porter-Thomas Statistics}

The following form \cite{Porter Thomas},%
\begin{equation}
P(I)=a\,\left(  \frac{I}{<I>}\right)  ^{b}\exp\left[  -c\left(  \frac{I}%
{<I>}\right)  \right]  \label{Porter}%
\end{equation}
was fit to the data, where $I$ is the peak current and $<I>$ the mean peak
current.\ In this analysis, parameters specific to the Porter-Thomas and
Poisson distributions, which were deemed relevant, were used.
\begin{figure}
[ptb]
\begin{center}
\includegraphics[
height=1.9173in,
width=3.2448in
]%
{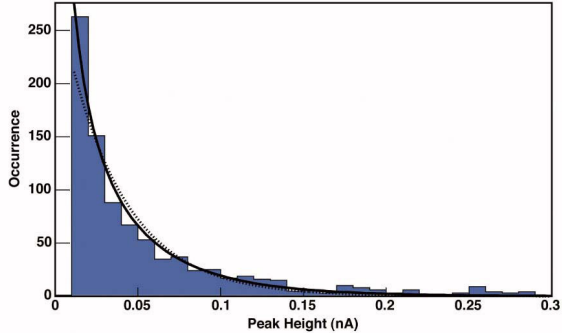}%
\caption{Histogram of peak heights for cluster no. 9. The solid curve is the
fit for the Porter-Thomas distribution. The dotted line represents the fit for
the Poisson distribution. There are 851 peaks heights that comprise the
histogram.}%
\label{PorterThomas}%
\end{center}
\end{figure}
The Porter-Thomas distribution (with $a=(2\pi)^{-\frac{1}{2}},\,b=-1/2$ and
$c=1/2)$ provided a somewhat better fit to the data than the Poisson
distribution. Figure \ref{PorterThomas} shows a plot of the histogram along
with curves associated with the best fits of the Porter-Thomas and Poisson
distributions. The results of this analysis support the interpretation that
these measurements are yielding spectroscopic information relating to the
energy levels.

\subsection{Crossover between Poisson-like and Wigner-Dyson Statistics}

\begin{figure}
[ptb]
\begin{center}
\includegraphics[
height=7.1831in,
width=2.5573in
]%
{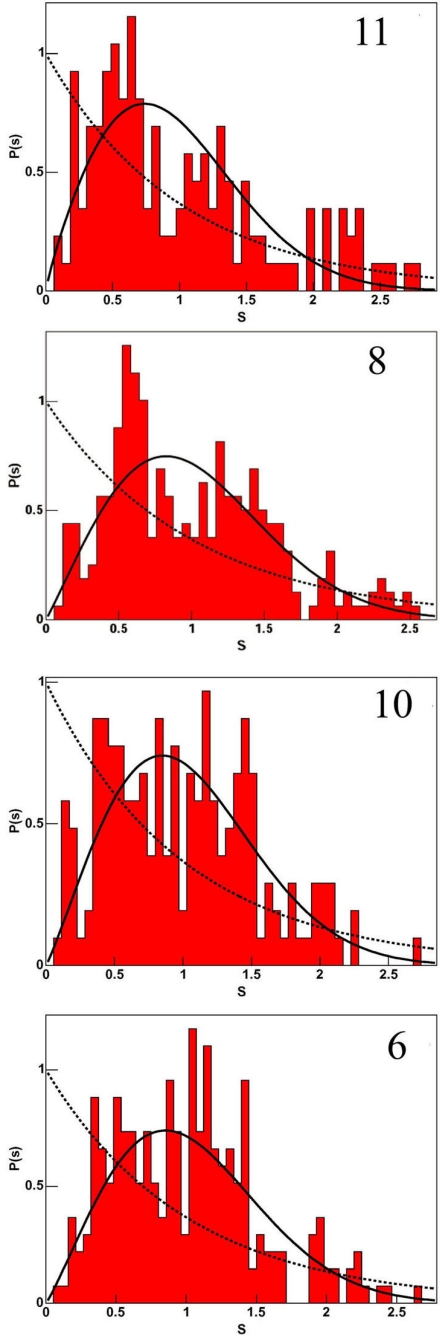}%
\caption{Transition to chaos in the level statistics of Pb clusters of the
same approximate volume ($\approx10$ $nm^{3})$ but with different shapes. The
solid black curve is the Wigner-Dyson distribution, while the dotted black
curve is the Poisson distribution for comparison. The cluster's catalog number
is in the upper right hand corner.}%
\label{transition}%
\end{center}
\end{figure}

Experimentally one should observe variations in the distributions of the
energy levels based on the cluster's shape. Keeping all the experimental
parameters the same, i.e. the same tip-cluster height distance and the same
voltage scale, the distributions of the eigenvalues change depending on the
cluster's shape. In the work presented here, the majority of distributions for
the different clusters fell into the regime between Wigner-Dyson and Poisson
statistics, which describes a partially chaotic system. This is plotted in
Fig. \ref{transition} where clusters of the same approximate volume but have
different shapes are considered. \ The top histogram in Fig. \ref{transition}
is more Poisson-like while the bottom histogram is strongly Wigner-Dyson
like.
This is in agreement with other experimental situations, such as the He atom
in a magnetic field \cite{Karremans} and the acoustic resonances of Al blocks
where deformations in the shape of the block away from parallelpiped generate
partially chaotic acoustic waves. \cite{Ellegaard} Completely chaotic systems
are rare and difficult to achieve experimentally.

The modified Berry-Robnik equation \cite{Podolskiy} indicates what percentage
of the distribution is Poisson like and what percentage is Wigner-Dyson like
and accounts for the small level repulsion that is observed in most
systems.(It is a modified version of the Berry-Robnik distribution
\cite{Berry} which works well for mixed chaotic and regular systems at the
tail of the distribution, but fails at small level spacings.). Thus the
modified Berry-Robnik equation is%

\begin{subequations}
\label{berryrobnik}%
\begin{align}
P(s)  &  \propto\,q^{2}\text{F(}\frac{s}{\nu^{2}})\text{e}^{-qs}%
\text{erfc}\left(  \frac{\sqrt{\pi}}{2}\left(  1-q\right)  s\right)  +\\
&  +\left[  \frac{\pi}{2}\left(  1-q\right)  ^{2}s+2qF\left(  \frac{s}{\nu
}\right)  \right]  \left(  1-q\right)  e^{-qs-\frac{\pi}{4}\left(  1-q\right)
^{2}s^{2}}%
\end{align}
where erfc is the complimentary of the error function and is defined as%

\end{subequations}
\begin{equation}
\text{erfc}\left(  x\right)  \text{=1-}\operatorname{erf}\left(  x\right)
=\frac{2}{\sqrt{\pi}}%
{\displaystyle\int\limits_{x}^{\infty}}
e^{-u^{2}}du \label{complimentaryerrorfunction}%
\end{equation}

In addition, $F(x)$ is defined as follows:%

\begin{equation}
F\left(  x\right)  =1-\frac{1-\sqrt{\frac{\pi}{2}}x}{e^{x}-x} \label{F(x)}%
\end{equation}
When q = 0 the Wigner-Dyson term is retrieved and when q=1, the distribution
is a Poisson distribution. Equation \ref{berryrobnik} describes a mixed state
between these two limits and addresses the crossover regime in which the
dynamics are partially chaotic.In Table \ref{Pb Parameters copy(1)}, the
fitted values for the modified Berry Robnik equations are summarized
for\textit{ different sized }Pb clusters.
\begin{table}[tbp] \centering
\begin{tabular}
[c]{|c|c|c|c|c|}\hline
Cluster & \# I-V & \# Peak & q & $\nu$\\
& curves & Spacings &  & \\\hline
1 &
$<$%
10 &
$<$%
30 & --- & -----\\\hline
2 & 180 & 35 & --- & -----\\\hline
3 & 119 & 137 & .59$\pm0.71$ & .65 $\pm\,0.22$\\\hline
4 & 137 & 144 & 0.02 $\pm\,0.06$ & ---\\\hline
5 & 125 & 16 & --- & ---\\\hline
6 & 158 & 254 & -3.1e-9 $\pm\,4.39e-02$ & ---\\\hline
7 &
$<$%
10 &
$<$%
30 & --- & ---\\\hline
8 & 207 & 273 & .14$\pm0.085$ & .45$\pm0.17$\\\hline
9 & 280 & 413 & 0.02$\pm0.037$ & ---\\\hline
10 & 131 & 181 & .06 $\pm.062$ & .18$\pm0.25$\\\hline
11 & 316 & 142 & 1.37 $\pm\,0.56$ & .73$\pm0.06$\\\hline
12 &
$<$%
10 &
$<$%
30 & --- & ---\\\hline
13 & 193 & 237 & 0.004$\pm0.05$ & ---\\\hline
14 & 405 & 64 & 1.28$\pm0.58$ & .7$\pm0.16$\\\hline
15 & 220 & 105 & -0.0086$\pm.07$ & ---\\\hline
16 & 263 & 450 & 0.25$\pm0.07$ & .48$\pm.08$\\\hline
\end{tabular}%
\caption{Summary of the statistics of different sized Pb clusters. When q=0 or
close to it, WD statistics prevails and the value of $\nu$ becomes
meaningless, and when q = 1 Poisson statistics is the appropriate description
of the distribution. (The low statistics for cluster numbers 1, 2, 5, 7,
and 12 resulted in undefined q and $\nu$ values.)}%
\label{Pb Parameters copy(1)}%
\end{table}%

\subsection{Absence of Charging Energy, Superconductivity and Spin-Orbit
Scattering}

An important issue is the role of charging energy in the proposed two-step
tunneling process. It is likely that the charging energy is absent because of
additional capacitance arising from the proximity of neighboring clusters.
This will lower the charging energy significantly so that the peaks that are
observed are only those related to the eigenstates of the clusters. \ Absence
of charging energy was also recently reported from x-ray photoemission
spectroscopy measurements \cite{Salmeron} on gold nanocrystals which were
self-assembled using wet chemistry. In this study, as the mutual separation
between gold clusters was reduced to distances less than 1nm by varying the
ligand size, the charging energy was eliminated.\ New experimental conditions
need to be realized in our work in which the clusters are well separated from
each other in order to resolve this issue about the charging energy.

The apparent absence of superconductivity is a result of the mean level
spacing exceeding the energy gap of superconducting Pb which is $\thicksim
2\,$\thinspace$meV$. For this reason one would not expect to observe features
in the tunneling characteristic associated with superconductivity.
\cite{Anderson} This was also observed in the pioneering work by Ralph, Black
and Tinkham. \cite{Ralph}

Lead is a strong spin-orbit coupling material and the appropriate description
to describe strong spin orbit coupling is a Gaussian Sympletic Ensemble (GSE)
instead of Gaussian Orthogonal Ensemble (GOE\ or Wigner-Dyson) distribution.
\ How strongly spin orbit coupling adheres to GSE\ instead of GOE can be
related to the cluster's size \cite{Matveev} as follows: \ If Wigner-Dyson
(GOE) statistics prevail, then the following criteria based on the
uncertainity principle ($\hbar=1)$ should be met: \ $\delta E\delta\tau
_{SO}\gg1.$ Likewise, if the appropriate statistics is related to Gaussian
Sympletic Ensemble (GSE), then the following criteria is applicable: $\delta
E\delta\tau_{SO}\ll1$ . In these relations, $\delta E$ is the mean energy
level spacing, or one divided by the density of states. \ The term $\delta
\tau_{SO}$ is the time that it takes to flip a spin or $\delta\tau_{SO}%
=\frac{L}{v_{f}}$ where $L$ is the length of the cluster, and $v_{f}$ is the
Fermi velocity. This assumes that the lateral transport in the dot is
ballistic with the only scattering occuring at the boundaries. Thus the
expression $\delta E\delta\tau_{SO}$ is equal to $\frac{2\pi^{2}L}{k_{f}%
^{2}\,Vol}.$ In the Table \ref{Pb Parameters} the values for $\delta
E\delta\tau_{SO}$ are calculated based on cluster size instead of the actual
mean level spacing which is smaller than theoretical estimates. This upper
limit indicates that the appropriate distribution should be GSE\ instead of
GOE. This contradicts our experimental findings.

It should be noted that in studies of Au grains (%
$<$%
5 nm), it was found that the spin-orbit scattering was suppressed.
\cite{Davidovic} It was speculated that the origin of the suppression of the
spin-orbit scattering was the granularity of the weakly coupled grains.
Estimates of this suppression were carried out by making weak localization
measurements. This involves a study of the orbital effect in an applied
magnetic field. \ While we could not make such estimates (our apparatus lacks
a magnetic field) using similar arguments, we may have observed a suppression
of spin-orbit scattering (based on the statistics).

\section{IMAGING QUANTUM CHAOS WITH AN STM}

\begin{figure}
[ptb]
\begin{center}
\includegraphics[
height=2.5313in,
width=3.1609in
]%
{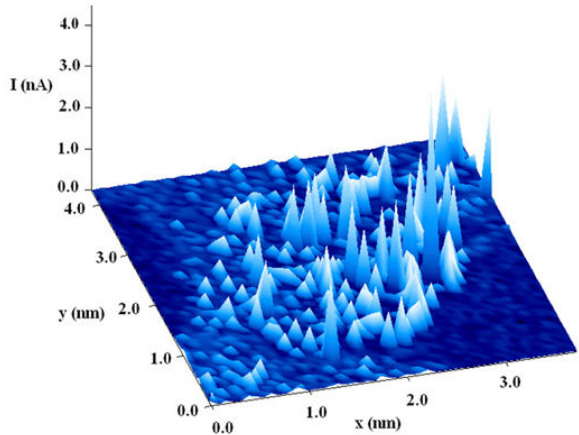}%
\caption{Current as a function of position on a chaotic cluster (cluster no.
4) at 26 meV.}%
\label{imagechaos}%
\end{center}
\end{figure}

Finally we turn our attention to imaging the eigenfunctions of the clusters
using a scanning tunneling microscope. The wavefunction moduli are directly
related to the tunneling current, so by mapping the tunneling current
amplitude at a fixed bias, it is possible to study the significance of chaos
as a function of cluster shape. An image of the current peaks is shown in Fig.
\ref{imagechaos} for cluster no. 4. \ The current peaks change as a function
of applied bias (not shown).%

The current dependence as a function of bias voltage can be used to
\textquotedblleft image\textquotedblright\ the chaotic nature of irregularly
shaped metallic clusters. This gained prominence in the experimental community
with the pioneering experiments of Sridhar and co-workers. \cite{Sridhar}
Besides acquiring images of the clusters, it is possible to produce images of
chaotic behavior in the same manner than was carried out by Sridhar using
electric field to probe the distributions of eigenfunctions in a
\textquotedblleft Sinai and rectangular\textquotedblright\ structure. In
essence scanning tunneling spectroscopy is a technique that can be used to map
out the amplitude of the eigenfunctions which are otherwise difficult to access.

\section{CONCLUDING REMARKS AND OPEN QUESTIONS}

The empirical evidence indicates that discrete energy levels are being
accessed from a resonant tunneling geometry and that the distributions of
these energy levels reflect the underlying classical dynamical nature of an
electron under confinement of its boundary. This means that our study of the
distribution of spacings gives information about the symmetry properties of
the physical system in question. In this study of metal clusters, there are
remaining questions that still need be resolved. \ First what is the nature of
the tunneling process? \ Is the electron's route through an Si interface
state, or through neighboring clusters? \ Future experimental work needs to be
carried out in which the clusters are separated from one another in order to
resolve this issue. The second question that needs to be answered is why are
the spin-orbit interactions suppressed? \ That is, why not Gaussian Sympletic
Ensemble statistics instead of Wigner- Dyson statistics? This issue needs to
be explored both theoretically and experimentally in greater detail. \ 

\textbf{ACKNOWLEDGMENTS}

It is a pleasure to thank the experts in this field that have provided useful
and stimulating discussions with us about our work. Specifically we thank
Leonid Glazman, Alex Kamenev, and Denis Ullmo. Also, one of us (L. L. A. A.)
would like to thank the organizers, instructors and participants attending the
Boulder Condensed Matter Summer School (2005) for many lively and helpful
discussions. This work was supported by the US\ Department of Energy under
grant DE-FG02-02ER46004.

\clearpage

\noindent

\end{document}